\documentclass[preprint,tightenlines,nobibnotes]{revtex4}

\usepackage{graphicx}

\begin{document}

\title{Constraining the equation of state of nuclear matter from competition of fusion and quasi-fission in the reactions leading 
to production of the superheavy elements}

\author{M. Veselsky}
\email{Martin.Veselsky@savba.sk}
\affiliation{Institute of Physics, Slovak Academy of Sciences, Bratislava, Slovakia}
\author{J. Klimo}
\affiliation{Institute of Physics, Slovak Academy of Sciences, Bratislava, Slovakia}
\author{Yu-Gang Ma}
\affiliation{Shanghai Institute of Applied Physics, Chinese Academy of Sciences,
Shanghai 201800, China}
\author{Georgios A. Souliotis}
\affiliation{Laboratory of Physical Chemistry, Department of Chemistry,
National and Kapodistrian University of Athens, Athens 15771, Greece}




\begin{abstract}
The mechanism of fusion hindrance, an effect preventing the synthesis of superheavy elements 
in the reactions of cold and hot fusion, is investigated using 
the Boltzmann-Uehling-Uhlenbeck equation, where Coulomb 
interaction is introduced. A strong sensitivity is observed both to 
the modulus of incompressibility of symmetric nuclear matter, 
controlling the competition of surface tension and Coulomb repulsion, 
and to the stiffness of the density-dependence of symmetry 
energy, influencing the formation of the neck prior to scission. The experimental 
fusion probabilities were for the first time used to derive constraints on the nuclear equation of state. 
A strict constraint on the modulus of incompressibility of nuclear matter $K_0 = 240 - 260$ MeV 
is obtained while the stiff density-dependences of the 
symmetry energy ($\gamma>1.$) are rejected. 
\end{abstract}

\maketitle


In the last two decades of the past century, the heavy elements up to Z=112 
were synthesized using cold fusion reactions 
with Pb, Bi targets in the evaporation channel with emission 
of one neutron \cite{Z112Cold}. The experimentalists 
had to face a rapid decrease of cross sections down to the picobarn 
level due to increasing fusion hindrance whose origin was unclear. 
Since the turn of millennium, still heavier elements with Z=113-118 were produced 
in the hot fusion reactions with emission of 3-4 neutrons using $^{48}$Ca beams 
with heavy actinide targets between uranium and californium \cite{SHECSSyst,SHECSSyst2,SHECSSyst3,SHECSSyst4,SHECSSyst5,SHECSSyst6,SHECSSyst7}.
Again the increase of fusion hindrance was observed, caused by competition of the fusion process 
with an alternative process called quasi-fission. 
Quasi-fission occurs when instead of fusion the 
system forms elongated shape evolves towards the scission point. 
The systematics in the reactions with lead target, 
published in \cite{Boc82}, shows that quasi-fission sets on for beams $^{48}$Ca and heavier. 
In terms of reaction mechanism, quasi-fission is similar to nucleon exchange between colliding nuclei, however 
it proceeds while the shape of the system also changes dramatically. Compared to the fusion-fission, proceeding via formation of 
compound nucleus, angular distribution of fission fragments is forward-peaked in center-of-mass frame, 
total kinetic energy (TKE) is lower and the mass asymmetry ranges from the mass asymmetry 
of projectile-target configuration towards the symmetric mass split, 
with the yield decreasing monotonously. A large systematics of high quality data 
on quasi-fission in the reactions, leading to production of superheavy elements, 
was obtained in the recent years in Dubna \cite{SHEItkis} and Tokai \cite{Nishio}. 
It is usually considered that the process of quasi-fission is governed by 
a complex dynamics of the projectile-target system, which is often described using 
theoretical tools such as the model of di-nuclear system \cite{SHEDiNuc} 
or the Langevin equation \cite{SHELang,Yoshi}. 
Besides the above theoretical tools, the competition of fusion
and quasi-fission was also addressed using the implementations 
of the Boltzmann equation known as ImQMD \cite{ImQMD} and 
using the time-dependent Hartree-Fock theory \cite{TDHF}. 
However, success of a simple statistical model of fusion hindrance, 
introduced in  \cite{MVSHE,SHEJadFiz} suggests that the 
competition of fusion and quasi-fission could be dominantly driven by the available 
phase-space and hindrances originating in diabatic dynamics are not decisive.  
In the present work we employ the Boltzmann-Uehling-Uhlenbeck (BUU) equation 
with the Pauli principle implemented separately for neutrons and protons 
and with the Coulomb interaction. 
We demonstrate how various equations of state 
of nuclear matter implemented into such transport simulation 
influence the competition of fusion 
and quasi-fission. Based on available data on reactions, 
leading to production of super-heavy nuclei, we extract 
most stringent constraints on the stiffness of the nuclear 
equation of state and on the density-dependence of the symmetry energy. 


In order to describe theoretically the competition of fusion and 
quasi-fission at energies close to the Coulomb barrier, the goal is  
to describe the evolution of the nuclear mean field of the two reaction 
partners. However, besides nuclear mean field it is necessary 
to take into account the electrostatic interaction among protons 
and also it is necessary to guarantee preservation of the Pauli 
principle in a strict way. 
The evolution of nuclear mean field can be described by solving the Boltzmann equation. 
One of the approximations for the solution of the Boltzmann equation, 
the Boltzmann-Uehling-Uhlenbeck model is extensively used 
\cite{BUU1,BUU2}, which takes both the nuclear mean field and the Fermionic Pauli
blocking into consideration. The BUU equation reads

\begin{eqnarray}
      & \frac{\partial f}{\partial t}+ v \cdot \nabla_r f - \nabla_r U
\cdot \nabla_p f  = \frac{4}{(2\pi)^3} \int d^3p_2 d^3p_3 d\Omega
\nonumber
\\ & \frac{d\sigma_{NN}}{d\Omega}v_{12}
 \times [f_3 f_4(1-f)(1-f_2) - f f_2(1-f_3)(1-f_4)] \nonumber
\\ & \delta^3(p+p_2-p_3-p_4),  \label{BUU}
                   \end{eqnarray}

\noindent
where  $f$=$f(r,p,t)$ is the phase-space distribution function. 
It is solved with the test particle method of Wong 
\cite{Wong}, with the collision term as introduced  
by Cugnon et al. \cite{Cugnon}. 
In Eq.(~\ref{BUU}), $\frac{d\sigma_{NN}}{d\Omega}$
and $v_{12}$ are in-medium nucleon-nucleon cross section and 
relative velocity for the colliding nucleons, respectively, and
$U$ is the sum of the simple single-particle mean field potential 
with the isospin-dependent symmetry energy term 

\begin{equation}
{U} =  a\rho + b\rho^{\kappa}
+ 2{a_s}(\frac{\rho}{\rho_0})^{\gamma} \tau_z I , 
\label{EqPot}
\end{equation}

\noindent
where 
$I=(\rho_n-\rho_p)/\rho$, 
$\rho_0$ is the normal nuclear matter density; $\rho$,
$\rho_n$, and $\rho_p$ are the nucleon, neutron and proton
densities, respectively; 
$\tau_z$ assumes the value 
1 for neutron and -1 for proton,  
the coefficients $a$, $b$ and exponent $\kappa$ represent the properties of  
symmetric nuclear matter, while the last term describes the influence 
of the symmetry energy,  
where $a_s$ represents 
the symmetry energy at saturation density  
and the exponent $\gamma$ describes the density dependence. 

The in-medium nucleon-nucleon cross sections are typically 
approximated using the experimental cross sections of free nucleons 
(e.g. using the parametrization from Cugnon \cite{Cugnon}). 
Alternatively, as shown in the work \cite{VdWBUU}, in-medium 
nucleon-nucleon cross sections can be 
estimated directly using the equation of state 
and used successfully e.g. to describe the evolution of transverse 
flow in a wide range of relativistic energies \cite{MVAPS,Flow}. 
However, at low energies close to the Coulomb barrier 
the collision term plays only limited role 
and the choice of the in-medium nucleon-nucleon cross sections 
does not influence the results of simulations, in part because 
at such low relative momenta both cross sections exceed the cutoff value 
applied in the BUU code. 

In order to describe the nuclear collisions close to 
the Coulomb barrier, it is crucial to implement properly 
the electrostatic interaction among protons. It is however 
impossible to introduce a density-dependent 
term into the single-particle potential in the equation (\ref{BUU}), 
since electrostatic interaction has a long-range 
and for the infinite nuclear matter it would diverge.  
In the present work, instead of modification of the 
single-particle potential, we complement the corresponding 
density-dependent force $\nabla_r U$ acting at a given cell of the 
cubic grid (with a mesh of 1 fm) with the 
summary force generated by the proton distribution outside 
of the cell. This approach thus avoids fluctuation of the 
Coulomb force due to interaction of protons inside the cell, 
what is natural since this interaction is considered in the 
collision term. Specifically we consider only interaction with the protons 
of the same set of test particles, which allows to perform 
simulations practically with the same CPU time consumption as the simulations 
without Coulomb interaction. This circumstance allowed to perform this study in principle.  
Besides the introduction of Coulomb interaction, at low beam energies 
close to the Coulomb barrier it is necessary to guarantee strict 
preservation of the Pauli principle. We assure this by implementing 
the Pauli principle separately for protons and neutrons. 

The simulations were performed using various assumptions 
on the stiffness of the equation of state of symmetric nuclear matter, 
as represented by the single-particle potential in Eq. (\ref{EqPot}). 
The exponent $\kappa$ was varied between values of $7/6$ and $2$, 
corresponding to the range of incompressibilities between 200 and 380 MeV 
(the value of incompressibility depends linearly on $\kappa$). 
Besides the stiffness of the equation of state of symmetric nuclear matter, 
we implemented several assumptions on the stiffness of the density dependence 
of symmetry energy by varying the exponent $\gamma$ in Eq. (\ref{EqPot}) 
between 0.5 and 1.5. For each calculated reaction, the simulation was 
performed using 600 test particles, with 20 different sequences of the 
pseudo-random numbers. 
The simulations were performed using a computing workstation 
with four Xeon Phi coprocessor cards with 61 cores, allowing 
to perform hundreds of simulations (up to one thousand) in parallel. 

\begin{figure}[t]
\centering
\includegraphics[height=8.5cm,width=13.5cm]{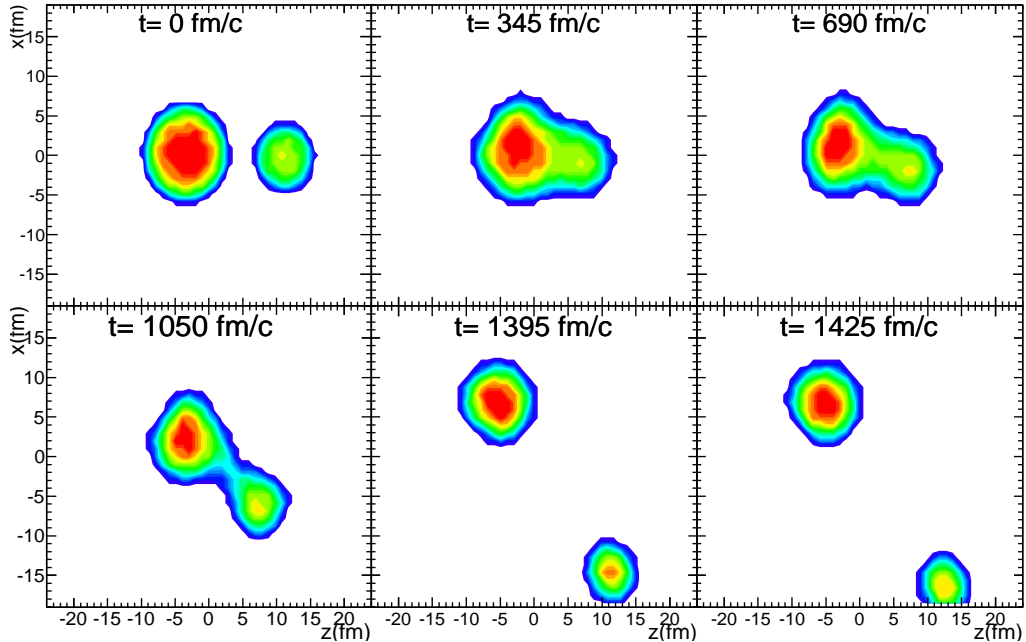}
\caption{ 
Typical evolution of nucleonic density for the central collision 
$^{64}$Ni+$^{186}$W at beam energy 5 AMeV, simulated using the soft equation of state with 
incompressibility $K_0 = 202$ MeV and the soft 
density dependence symmetry energy with $\gamma = 0.5$. 
Weak surface tension is overcome by Coulomb interaction 
and quasi-fission occurs. 
}
\label{figk202g05}
\end{figure}


In order to investigate the role of the equation of state of nuclear matter 
in the competition of fusion and quasi-fission in reactions leading to 
heavy and superheavy nuclei, we selected a representative set of reactions, 
where experimental data exists. As one of the heaviest systems, where 
fusion is still dominant, we use the reaction $^{48}$Ca+$^{208}$Pb. This 
reaction was measured \cite{ExpNiPb,ExpCaPb}, and a typical dominant peak at 
symmetric fission was observed in the mass vs TKE spectra of fission fragments, 
with TKE consistent to fusion-fission proceeding through 
formation of the compound nucleus $^{256}$Nb. Onset of quasi-fission was observed \cite{ExpNiW} 
in the reaction $^{64}$Ni+$^{186}$W, leading to compound system $^{250}$No, where 
a prominent fusion-like peak is not observed anymore, however 
symmetric fission, which can be attributed to fusion-fission, is still 
observed relatively frequently. 
Quasi-fission becomes even more dominant in the reaction $^{48}$Ca+$^{238}$U, 
nominally leading to compound nucleus $^{286}$Cn. Nevertheless, the 
symmetric fission events still amount to about 10 \% of fission events \cite{ExpCaU}. 
Comparison with the reaction  $^{64}$Ni+$^{186}$W shows that the relative amount of symmetric events 
in reaction $^{48}$Ca+$^{238}$U is twice lower than in the reaction $^{64}$Ni+$^{186}$W, 
thus implying the relative amount of 20 \% of symmetric fission for the latter reaction.  
In reactions $^{64}$Ni+$^{208}$Pb \cite{ExpNiPb}, $^{48}$Ca+$^{249}$Cf \cite{SHECSSyst6}, 
and $^{64}$Ni+$^{238}$U \cite{ExpNiU} the quasi-fission already dominates and fusion hindrance 
amounts to several orders of magnitude (10$^{-3}$ - 10$^{-5}$ \cite{MVSHE,SHEJadFiz}). 

\begin{figure}[t]
\centering
\includegraphics[height=8.5cm,width=13.5cm]{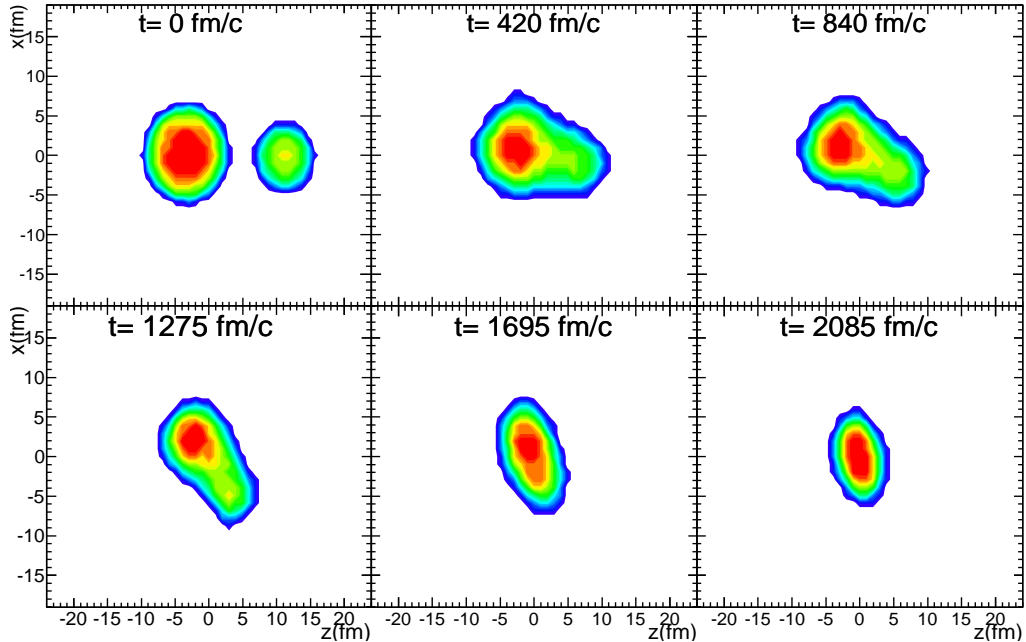}
\caption{ 
Typical evolution of nucleonic density for the central collision 
$^{64}$Ni+$^{186}$W at beam energy 5 AMeV, simulated using the stiff equation of state ($K_0 = 300$ MeV) and the soft 
density-dependence of symmetry energy ($\gamma = 0.5$). 
Stronger surface tension overcomes Coulomb interaction 
and quasi-fission is prevented. 
}
\label{figk300g05}
\end{figure}

Simulations were performed at beam energy 5 AMeV, which is above the Coulomb barrier and in all cases corresponds to the 
nearest experimental point within few MeV. 
Since the angular momentum range where quasi-fission events are produced 
is not known precisely and also to assure that we won't observe deep-inelastic transfer, 
which occurs at peripheral collisions, we simulate the 
most central events, with impact parameter set to 0.5 fm (exactly central 
events practically do not occur in experiment).  
Simulations were performed up to the time 3000 fm/c, 
sufficient for formation of the final configuration in all 
investigated cases. 
To carry out comparison with experiment, we need to determine from available experimental 
information the probability of fusion in central collisions. 
For the reaction $^{48}$Ca+$^{208}$Pb the fusion probability is  
close to 100 \%, while for reactions $^{64}$Ni+$^{208}$Pb, $^{48}$Ca+$^{249}$Cf, 
and $^{64}$Ni+$^{238}$U it is close to zero (10$^{-3}$ - 10$^{-5}$). 
Of the two remaining reactions, the total fusion probability of 10 \% 
and the fact that fusion probability peaks at central collisions 
infer the constraint on fusion probability in the reaction $^{48}$Ca+$^{238}$U 
at central events between 20 - 50 \% 
(upper limit is based on assumption that quasi-fission is dominant even in central collisions). 
Since comparison of shapes of experimental mass distribution in reactions $^{48}$Ca+$^{238}$U 
and $^{64}$Ni+$^{186}$W shows that there is approximately   
twice higher relative abundance of fusion in reaction $^{64}$Ni+$^{186}$W, 
we constrain the fusion probability in this reaction 
at central collisions between 40 - 80 \%. 
These constraints remain still relatively loose, but 
they reflect the dominant scenarios and thus 
the dynamics of competition of fusion and quasi-fission. 
As a consequence, this representative set of reactions allows to constrain 
the parameters of equation of state of nuclear matter. 

\begin{figure}[t]
\centering
\includegraphics[height=8.5cm,width=13.5cm]{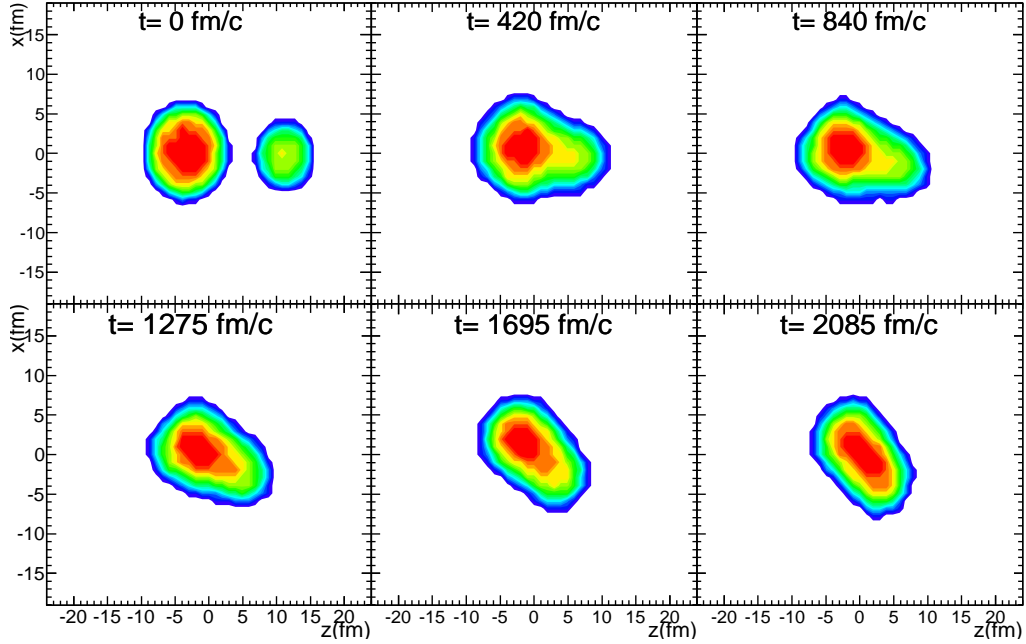}
\caption{ 
Typical evolution of nucleonic density for the central collision 
$^{64}$Ni+$^{186}$W at beam energy 5 AMeV, simulated using the soft equation of state ($K_0 = 202$ MeV) 
and the stiff density-dependence of symmetry energy ($\gamma = 1.5$). 
Despite weak surface tension the stiff 
density-dependence of symmetry energy prevents formation 
of a neck and quasi-fission is prevented. 
}
\label{figk202g15}
\end{figure}

From the investigated reactions, the collisions of $^{64}$Ni+$^{186}$W exhibit 
highest sensitivity to the parameters of the equation of state. 
Fig. \ref{figk202g05} shows typical evolution of nucleonic density for the collision 
$^{64}$Ni+$^{186}$W, simulated using the soft equation of state with 
incompressibility $K_0 = 202$ MeV and the soft 
density-dependence of symmetry energy $\gamma = 0.5$. 
One can see that the impinging projectile nucleus establishes contact 
with the target nucleus, however the weak surface tension, caused by 
the soft equation of state, is not sufficient to overcome Coulomb repulsion 
of the projectile and target which re-separate after approximately 
1200 fm/c (scission time is comparable with other approaches \cite{SHEDiNuc,SHELang,Yoshi,ImQMD,TDHF}). 
Similar evolution was observed in all 20 simulated test particle sets. 
Fig. \ref{figk300g05} 
shows simulation of the same reaction with $K_0 = 300$ MeV and $\gamma = 0.5$. 
In all simulations of this case the stronger surface tension generated by the stiff equation 
of state allows to overcome the Coulomb repulsion and system undergoes fusion. 
Strong sensitivity to the stiffness of the equation of state is thus demonstrated. 
Fig. \ref{figk202g15} shows the simulation with $K_0 = 202$ MeV and  $\gamma = 1.5$. One can 
observe that increased stiffness of the density-dependence of symmetry 
energy can also prevent system from separating into two fragments. In this case 
the weak surface tension allows to form elongated configuration (similar to  
Fig. \ref{figk202g05}), however the stiffer symmetry energy prevents formation 
of a low-density neutron-rich neck and thus the contact between the two reaction partners  
is preserved until the surface tension finally overcomes the Coulomb 
repulsion. Figs. \ref{figk202g05} - \ref{figk202g15} demonstrate a strong sensitivity 
of the system $^{64}$Ni+$^{186}$W to the parameters of the equation of state.  

\begin{figure}[t]
\centering
\includegraphics[height=9cm,width=10.5cm]{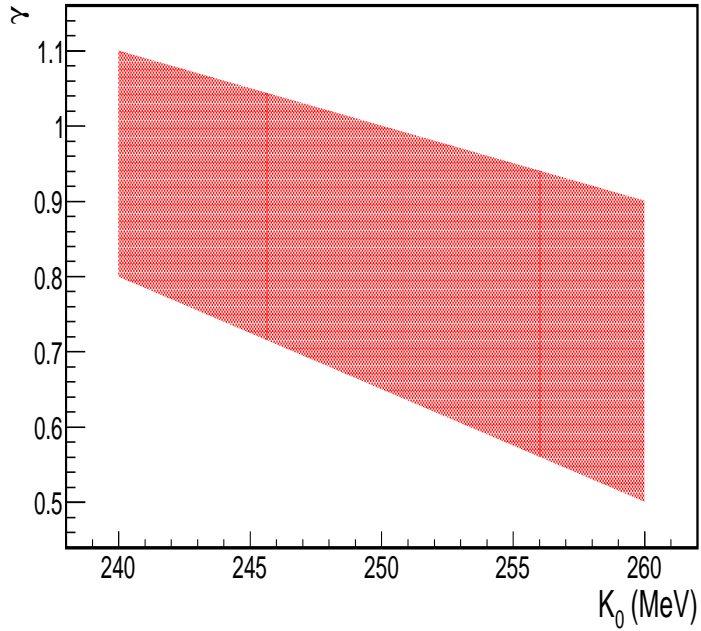}
\caption{ 
Constraint on stiffness of symmetric nuclear matter (modulus of incompressibility) and 
on density-dependence of the symmetry energy (exponent $\gamma$ from Eq. \ref{EqPot}) derived from the 
simulations of competition between fusion and quasi-fission. 
}
\label{figk0esym}
\end{figure}

In the other system of comparable mass, collisions of $^{48}$Ca+$^{208}$Pb typically result 
in fusion, with exception of the simulations with $K_0 = 202 - 230$ MeV 
and $\gamma = 0.5 - 1.0$. For such soft equation of state collisions 
usually result in quasi-fission and thus such equations of state can 
be considered in conflict with experiment. 
Heavier systems $^{64}$Ni+$^{208}$Pb, $^{48}$Ca+$^{249}$Cf,
and $^{64}$Ni+$^{238}$U usually undergo quasi-fission for 
$K_0 = 202 - 255$ MeV, for stiffer equations of state fusion 
appears and eventually becomes dominant. Thus a stiff equation of state with $K_0 = 272 - 300$ MeV 
can be rejected. Also a stiff symmetry energy with $\gamma = 1.5$ 
combined with soft equations of state with $K_0 = 202 - 255$ MeV lead to fusion and thus 
can be rejected. 
The remaining system $^{48}$Ca+$^{238}$U behaves similarly to $^{64}$Ni+$^{186}$W, 
consistently with constraints derived from other systems. 

As a result of the analysis of competition between fusion and quasi-fission, 
it was possible to set a rather strict 
constraint on the incompressibility of the equation of state of nuclear 
matter $K_0 = 240 - 260$ MeV  with softer density dependence of the symmetry 
energy with $\gamma = 0.5 - 1.0$ (see Fig. \ref{figk0esym}). 
This constraint is based on simulations of collisions, where maximum density reaches 1.4 - 1.5 times the 
saturation density. 
The shape of the constrained area reflects a trend of softening the density-dependence of 
the symmetry energy, necessary to balance increase of incompressibility. 
Such trend stems from competition of the surface 
tension, related to the stiffness of the equation of state of symmetric nuclear matter, 
with the Coulomb repulsion. This corresponds to the traditional picture 
of nuclear fission, where fissility of the system is defined as a ratio 
of the Coulomb repulsion to twice the surface energy. However, the present 
analysis goes beyond this simple macroscopic picture and elucidates the 
crucial role of the density-dependence of the symmetry energy in the 
dynamics of the system close to the scission point. In comparison with 
other methods, such as constraining the equation of state using the 
nuclear giant resonances \cite{Youn99,Stone,Piek} or the flow observables in relativistic nucleus-nucleus collisions 
\cite{BALiPhRep}, in the present analysis the effect of nuclear equation of state is manifested directly, 
and thus it is not affected by uncertainty related e.g. to description of underlying 
nuclear structure in the former or disentangling the effect 
of the two-body dissipation via nucleon-nucleon collisions in the latter case. 
In our simulations the microscopic nuclear shell structure is not considered. 
Role of the shell structure in nuclear fission remains an open question, as 
demonstrated by the recent observation of asymmetric fission of $^{180}$Hg \cite{Fiss180Hg}, 
contrary to expectations, based on the shell structure of fission fragments. 
The effect of closed nuclear shells can be manifested differently  
in fusion and quasi-fission channel, specifically for each system,  
and thus no simple trend must necessarily 
exist. By using a representative set of investigated reactions we provide a solid base 
for assumption that the extracted constraints do not depend critically 
on the effect of shell structure. In order to obtain even more strict constraints, in particular  
on the density-dependence of symmetry energy, more experimental data are necessary close 
to the onset of quasi-fission, where the sensitivity of the neck dynamics to 
the density dependence of the symmetry energy is highest. 
 
This work is supported by the Slovak Scientific Grant Agency under contract 
2/0121/14, by the Slovak Research and Development Agency under contract
APVV-0177-11 (M.V.), in part by the Major State Basic Research
Development Program in China under Contract No. 2014CB845400, the National
Natural Science Foundation of China under contract Nos. 11421505 and
11520101004 (Y.G.M.) and 
by ELKE account No 70/4/11395 of the National and Kapodistrian University of
Athens (G.S.).

\end{document}